# Maximizing the performance for microcomb-based microwave photonic transversal signal processors

*Yang Sun, Jiayang Wu, Senior Member, IEEE, Yang Li, Xingyuan Xu, Guanghui Ren, Mengxi Tan,*
*Sai Tak Chu, Brent E. Little, Roberto Morandotti, Fellow, IEEE, Fellow, Optica,*
*Arnan Mitchell, Fellow, Optica, and David J. Moss, Life Fellow, IEEE, Fellow, Optica*

*Abstract*—Microwave photonic (MWP) transversal signal processors offer a compelling solution for realizing versatile high-speed information processing by combining the advantages of reconfigurable electrical digital signal processing and high-bandwidth photonic processing. With the capability of generating a number of discrete wavelengths from micro-scale resonators, optical microcombs are powerful multi-wavelength sources for implementing MWP transversal signal processors with significantly reduced size, power consumption, and complexity. By using microcomb-based MWP transversal signal processors, a diverse range of signal processing functions have been demonstrated recently. In this paper, we provide a detailed analysis for the processing inaccuracy that is induced by the imperfect response of experimental components. First, we investigate the errors arising from different sources including imperfections in the microcombs, the chirp of electro-optic modulators, chromatic dispersion of the dispersive module, shaping errors of the optical spectral shapers, and noise of the photodetector. Next, we provide a global picture quantifying the impact of different error sources on the overall system performance. Finally, we introduce feedback control to compensate the errors caused by experimental imperfections and achieve significantly improved accuracy. These results provide a guide for optimizing the accuracy of microcomb-based MWP transversal signal processors.

## I. INTRODUCTION

Ever-increasing data capacity in the information age is driving the demand for high-speed information processing. In contrast to conventional microwave signal processing based on electronics, that faces intrinsic bandwidth bottlenecks [1, 2], the use of photonic hardware and technologies to process high-bandwidth microwave signals, or microwave photonic (MWP) processing, can provide speeds orders of magnitude faster [3, 4], which is critical for high-speed processing applications [3-6].

In the past two decades, a range of high speed MWP processors have been demonstrated by employing different optical approaches, in both discrete and integrated form, as optical filtering modules to process microwave signals modulated on a single optical carrier [3, 7-16]. While successful, featuring high performance with dynamic tuning, these approaches provided only single processing functions with limited reconfigurability and fixed parameters. In contrast, MWP transversal signal processors, where the microwave signal is modulated onto multiple optical carriers with adjustable delays and weights before summing via photodetection [17, 18], have significant advantages in achieving highly reconfigurable processing [17, 18].

For MWP transversal signal processors, a large number of optical carriers forming discrete taps to sample the input microwave signal are needed to achieve a high accuracy. Despite the use of conventional multi-wavelength sources, such as discrete laser arrays [19-21] and fibre Bragg grating arrays [22-24], to offer the discrete taps, the numbers of available taps they can provide are normally restricted to be less than 10 – mainly due to the dramatic increase of the system size, power consumption, and complexity with the tap number. Recent advances in optical microcombs [25, 26] provide an effective way to circumvent such problem by generating a large number of wavelengths equally spaced by

Manuscript received XX X, XX; revised XX X, XX; accepted XX X, XX. Date of publication XX X, XX; date of current version XX X, XX. This work was supported in part by the Australian Research Council Centre of Excellence Project in Optical Microcombs for Breakthrough Science (No. CE230100006), the Australian Research Council Discovery Projects (No. DP150104327, DP190102773, and DP190101576), and the Swinburne ECR-SUPRA program. (*Corresponding authors: Jiayang Wu; David J. Moss.*)

Yang Sun, Jiayang Wu, Yang Li, and David J. Moss are with the Optical Sciences Center, Swinburne University of Technology, Hawthorn, VIC 3122, Australia (e-mail: yangsun@swin.edu.au; jiayangwu@swin.edu.au; yangli@swin.edu.au; dmoss@swin.edu.au).

Guanghui Ren, Mengxi Tan and Arnan Mitchell are with Integrated Photonics and Applications Centre, School of Engineering, RMIT University, Melbourne, 3000 VIC, Australia (e-mail: guanghui.ren@rmit.edu.au, mengxitan@rmit.edu.au, arnan.mitchell@rmit.edu.au).

Xingyuan Xu is with State Key Laboratory of Information Photonics and Optical Communications, Beijing University of Posts and Telecommunications, Beijing 100876, China (e-mail: xingyuanxu@bupt.edu.cn).

S. T. Chu is with the Department of Physics, City University of Hong Kong, Hong Kong (e-mail: saitchu@cityu.edu.hk).

B. E. Little is with QXP Technology Inc., Xi'an, China (e-mail: brent.little@qxptech.com).

Roberto Morandotti is with the INRS – Énergie, Matériaux et Télécommunications, Varennes, QC J3X 1S2, Canada (e-mail: morandotti@emt.inrs.ca).

Color versions of one or more of the figures in this letter are available online at http://ieeexplore.ieee.org.

Digital Object Identifier X

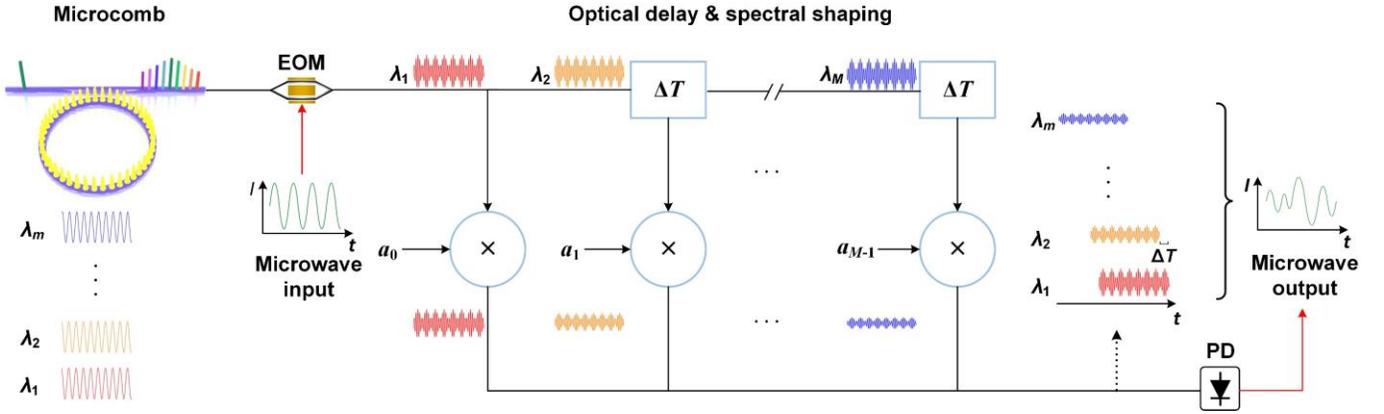

Fig. 1. Schematic diagram and signal processing flow of a MWP transversal signal processor with an optical microcomb source. EOM: electro-optic modulator. PD: photodetector.

large microwave bandwidths from single chip-scale devices. This opens new horizons for implementing MWP transversal signal processors with significantly reduced size, power consumption, and complexity. By using microcomb-based MWP transversal signal processors, a range of signal processing functions have been demonstrated recently, first for basic functions including differentiations [27, 28], integration [29], and Hilbert transforms [30-32], followed by more complex functions such as phase encoding [33], arbitrary waveform generation [34], and computations within the framework of optical neural networks [35-37].

For signal processors, processing accuracy is a key parameter. For microcomb-based MWP signal processors, processing errors are induced by both theoretical limitations and imperfect response of practical components. Recently, we presented an analysis quantifying the errors induced by theoretical limitations [38]. In this paper, we provide a complementary analysis to that work, focusing on errors induced by experimental imperfections. First, errors arising from imperfect microcomb characteristics, chirp in the electro-optic modulator, chromatic dispersion in the dispersive module, shaping errors of the spectral shaper, and noise of the photodetector are investigated. Next, a global picture is presented to show the influence of different error sources by quantifying their contributions to the overall system performance. Finally, we introduce feedback control to compensate errors induced by imperfect response of experimental components, and in doing so we achieve a significant improvement in the processing accuracy. These results are useful for understanding and optimizing the accuracy of microcomb-based MWP transversal signal processors.

## II. MICROCOMB-BASED MWP TRANSVERSAL SIGNAL PROCESSORS

Microwave transversal signal processors are implemented based on the transversal filter structure in digital signal processing that features a finite impulse response [37]. Implementing them with photonic technologies yields a significantly increased processing bandwidth compared to their electronic counterparts [17]. Fig. 1 shows the schematic diagram and signal processing flow of a typical MWP transversal signal processor. An optical microcomb, serving as a multi-wavelength source, provides a large number of wavelength channels as discrete taps. An input microwave signal is multicast onto each channel via an electro-optic modulator (EOM) to generate multiple microwave signal replicas. Next, time delays between adjacent wavelength channels are introduced by optical delay elements, and the delayed replicas at different wavelength channels are weighted through spectral shaping. Finally, the delayed and weighted replicas are summed via photodetection to generate the final microwave output of the system.

For the MWP transversal signal processor in Fig. 1, each of the taps can be regarded as a discrete sample of the system's impulse response, $i.e.$, the system's impulse response can be expressed as [17]

$$H(t) = \sum_{n=0}^{M-1} a_n \delta(t - n\Delta T), \qquad (1)$$

where $M$ is the tap number, $a_n$ ($n$ = 0, 1, 2, …, $M$-1) is the tap weight of the $n^{th}$ tap, and $\Delta T$ is the time delay between adjacent wavelength channels. Therefore, the output microwave signal $s(t)$ can be given by [39]

$$s(t) = f(t) * h(t) = \sum_{n=0}^{M-1} a_n f(t - n\Delta T), \qquad (2)$$

where $f(t)$ is the input microwave signal. After Fourier transformation from Eq. (1), the spectral transfer function of the MWP transversal signal processor is

$$H(\omega) = \sum_{n=0}^{M-1} a_n e^{-j\omega n\Delta T}, \qquad (3)$$

which shows agreement with the spectral response of a typical microwave transversal filter [39].

As can be seen from Eqs. (1) – (3), by simply altering the tap weights $a_n$ ($n$ = 0, 1, 2, …, $M$-1) through comb shaping, different signal processing functions can be achieved without any changes of the hardware [17]. This allows for a high

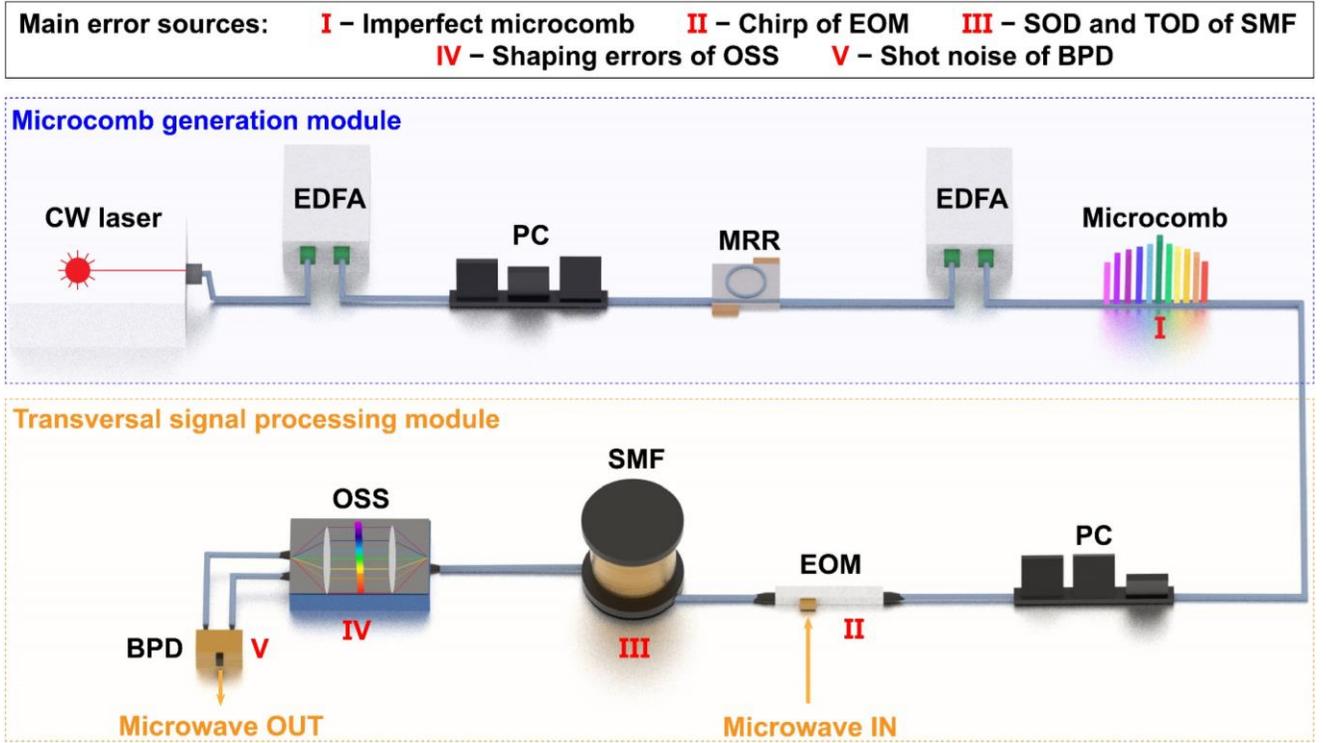

Fig. 2. Schematic of a practical microcomb-based MWP transversal signal processor. The main error sources are labelled as I – V. CW laser: continuous-wave laser. EDFA: erbium-doped fibre amplifier. PC: polarization controller. MRR: microring resonator. EOM: electro-optic modulator. SMF: single-mode fibre. OSS: optical spectral shaper. BPD: balanced photodetector. SOD: second-order dispersion. TOD: third-order dispersion.

degree of reconfigurability for the MWP transversal signal processor.

Fig. 2 shows a schematic of the experimental implementation of the MWP transversal signal processor in Fig. 1, which includes a microcomb generation module and a transversal signal processing module. In the microcomb generation module, a continuous-wave (CW) laser, amplified by an erbium-doped fibre amplifier (EDFA) with a polarization controller (PC) to adjust its polarization, is used to pump a high-Q nonlinear microring resonator (MRR) to generate optical microcombs. The output from this module is sent to the transversal signal processing module, which executes the signal processing flow depicted in Fig. 1. The processing module involves a PC, an EOM, a spool of single-mode fibre (SMF) as the optical delay module, an optical spectral shaper (OSS) to shape the comb lines, and a balanced photodetector (BPD) for photodetection. The BPD connected to the two complementary output ports of the OSS divides all the wavelength channels into two groups with a phase difference of $\pi$, which introduces positive and negative signs onto the tap coefficients $a_n$ ($n$ = 0, 1, 2, …, $M$-1) in Eqs. (1) – (3). It is worth noting that the particular processing function is determined not only by the absolute values of the tap coefficients but also by their signs. As an example, temporal integration is realized when all tap coefficients are set to 1 [29], whereas phase encoding can be achieved through the adjustment of specific coefficients to -1, while retaining the others at 1 [33].

For experimentally implemented MWP transversal signal processor in Fig. 2, processing errors arise from both theoretical limitations and imperfect response of practical system. The former refers to the theoretical approximation of a continuous impulse response (which corresponds to infinite tap number $M$) using a practical system with a finite tap number, and was the subject of our previous paper mentioned above [38]. The latter refers to errors induced by imperfect performance of different components, such as the noise of microcomb, chirp of the EOM, second- (SOD) and third-order dispersion (TOD) of the SMF, shaping errors of the OSS, and noise in the BPD.

To quantify the processing errors, the root mean square error (RMSE) is used to compare the deviation between the processor's output and the ideal result, which is defined as [40].

$$\text{RMSE} = \sqrt{\sum_{i=1}^{k} \frac{(Y_I - y_i)^2}{k}} \qquad (4)$$

where $k$ is the number of sampled points, $Y_1$, $Y_2$, …, $Y_n$ are the values of the ideal processing result, and $y_1$, $y_2$, …, $y_n$ are the values of the output of the microcomb-based MWP transversal signal processors.

Fig. 3(a) shows the RMSEs induced by theoretical limitations as a function of tap number $M$ for three different signal processing functions, including first-order differentiation (DIF), integration (INT), and Hilbert transform (HT). The theoretical limitations were calculated based on

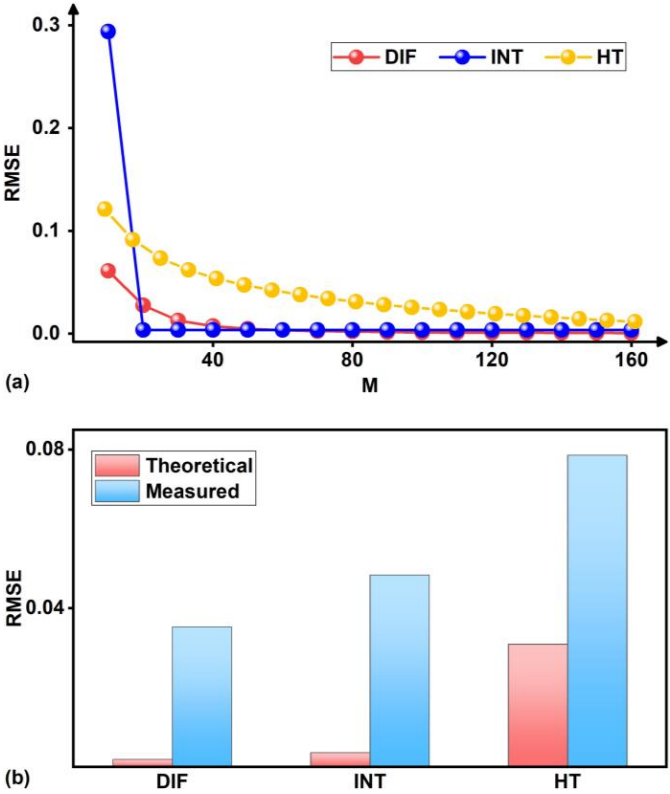

Fig. 3. (a) Root mean square errors (RMSEs) induced by theoretical limitation for differentiation (DIF), integration (INT), and Hilbert transformation (HT) as a function of tap number $M$. (b) Comparison of RMSEs induced by theoretical limitations and practical measured RMSEs for DIF, INT, and HT when $M = 80$. In (a) – (b), the comb spacing, length of dispersive medium, and second-order dispersion (SOD) parameter are $\Delta\lambda = 0.4$ nm, $L = 4.8$ km, and $D_2 = 17.4$ ps/nm/km, respectively. The input microwave signals are assumed to be Gaussian pulses with a full width at half maximum (FWHM) of ~0.17 ns.

Eqs. (1) – (4), assuming a perfect response for all the experimental components in Fig. 2. More details about this can be found in Ref. [41]. These theoretical RMSEs were calculated assuming a perfect response for all the components in Fig. 2. As can be seen, the theoretical RMSEs are small for a large tap number $M \geq 80$, indicating that the theoretical errors can be greatly reduced by increasing the tap number. Fig. 3(b) compares the theoretical and experimentally measured RMSEs for $M = 80$, showing that the former is much lower, reflecting that experimental errors typically dominate the system performance of microcomb-based MWP transversal signal processors. In the following Section III, we provide a comprehensive analysis of the experimentally induced processing errors, and in Section IV we provide approaches to mitigate these errors.

## III. ERRORS INDUCED BY IMPERFECTIONS OF PRACTICAL SYSTEMS

In this section, we provide a detailed analysis of the processing errors induced by different sources outlined in Fig. 2. This is achieved by modeling the imperfect response of the experimental components to calculate the output waveforms based on Eqs. (1) – (4). In subsections A – D, we investigate the influence of specific error sources, assuming the other sources are error-free. In subsection E, we compare the contributions of the different error sources to the overall system performance.

In the following analysis, we use first-order DIF, INT, and HT as examples to quantify the experimentally induced errors. Their spectral transfer functions are given by [27, 29, 31]

$$H_{DIF}(\omega) = j\omega, \quad (5)$$

$$H_{INT}(\omega) = \frac{1}{j\omega}, \quad (6)$$

$$H_{HT}(\omega) = \begin{cases} e^{-j\pi/2}, & 0 \leq \omega < \pi \\ e^{j\pi/2}, & -\pi \leq \omega < 0 \end{cases} \quad (7)$$

where $j = \sqrt{-1}$ and $\omega$ is the angular frequency.

For comparison, in our analysis we assume the processors have the same tap number ($M = 80$), comb spacing ($\Delta\lambda = 0.4$ nm), and length and SOD for the SMF ($L = 4.8$ km and $D_2 = 17.4$ ps/nm/km). These parameters are the same as those in our previous papers [27, 29, 31]. The input microwave signal is taken as a Gaussian pulse with a full width at half maximum (FWHM) of ~0.17 ns, whose spectral bandwidth (~5 GHz) is within the processing bandwidth of the signal processors (i.e., $FSR_{MW} = 1 / (\Delta\lambda \times L \times D_2) = $ ~30 GHz). For microcomb-based MWP transversal signal processors, the processing bandwidth is $min\{\Delta\lambda/2, FSR_{MW}/2\}$, where $min\{\cdot\}$ represents taking the minimum value between the two. Detailed elaboration on this can be found in Refs. [17, 18].

### A. Influence of the optical microcombs

In this section, we analyze the influence of microcomb imperfections on the system performance for different processing functions. These imperfections generate intensity and phase noise in the comb channels. The intensity noise includes power fluctuations of the comb lines and the intensity noise floor, which mainly arise from photon shot noise and spontaneous emission beat noise [42]. For MWP transversal signal processors, the microcomb intensity noise results in inaccuracy of the tap coefficients, thereby degrading the system accuracy.

To characterize the microcomb intensity noise, the optical signal-to-noise ratio (OSNR) is introduced, which is the ratio of the maximum optical signal to the noise power in each of the comb lines. Fig. 4(a) shows the simulated output waveforms from processors that perform DIF, INT, and HT, where flat intensity noise floors are assumed for the microcombs with different OSNRs. For comparison, the ideal processing outcome without theoretical errors, and the results that only account for theoretical errors (corresponding to OSNR = ∞) are also shown. As the OSNR of the comb lines increases from 10 dB to ∞, the processors' output waveforms match the ideal results better for all three processing functions, reflecting the reduced error achieved by increasing the OSNR. To better reflect the intensity envelop of the microcombs, a sinc-shaped intensity noise floor is introduced. The corresponding results are shown in Fig. 4(b), showing a trend similar to that in Fig. 4(a).

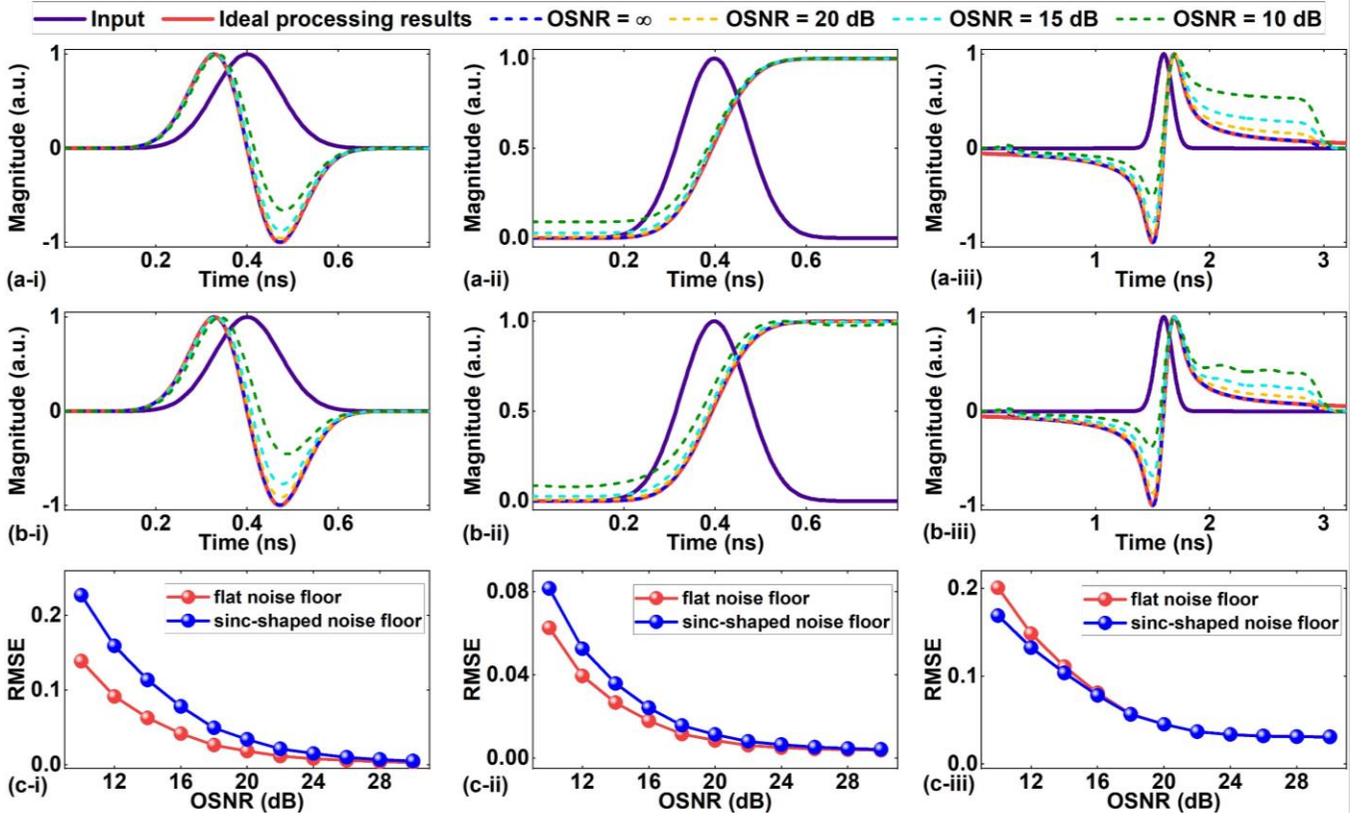

Fig. 4. Influence of microcombs' intensity noise on errors of differentiation (DIF), integration (INT), and Hilbert transformation (HT). (a) – (b) Temporal waveform of Gaussian input pulse and output waveforms from the transversal signal processors performing (i) DIF, (ii) INT, and (iii) HT, where the intensity noise floors of the microcombs are (a) flat and (b) sinc-shaped, respectively. Different curves show the results for different optical signal-to-noise ratios (OSNRs) of the comb lines. The ideal processing results are also shown for comparison. (c) Corresponding RMSEs between the ideal results and the processors' output waveforms as a function of microcomb's OSNR. In (a) – (c), the Gaussian input pulse has a FWHM of ~0.17 ns. The tap number, comb spacing, length of dispersive medium, and SOD parameter are $M = 80$, $\Delta\lambda = 0.4$ nm, $L = 4.8$ km, and $D_2 = 17.4$ ps/nm/km, respectively.

Fig. 4(c) shows the RMSEs between the simulated processors' output waveforms and the ideal processing results as a function of the OSNR. As expected, for both the flat and sinc-shaped intensity noise floor, the RMSEs decrease with the microcomb OSNR for all three processing functions, showing agreement with the trend in Figs. 4(a) and (b). For OSNRs less than 20 dB, the RMSEs decrease more steeply. As the OSNR increases, the decrease in RMSE is more gradual, and there is only a very small reduction in error beyond an OSNR of 20 dB. For the DIF and INT, the RMSE for microcombs with sinc-shaped intensity noise floors is higher than for flat intensity noise floors, whereas the opposite trend is observed for the HT. This reflects the fact that the impact of the microcomb intensity envelope errors depends on the processing function.

The phase noise of microcombs, which manifests as a broadened linewidth, an appearance of multiple repetition-rate beat notes, and a reduction in temporal coherence [43], is affected by several factors, such as the noise of the CW pump as well as the mechanical and thermal noise of the MRR [44, 45]. These sources of error are difficult to quantitatively analyze. For mode-locked microcombs with extremely low phase noise, the phase noise induced errors are negligible [35, 36]. Therefore, to achieve a high accuracy over long periods, it is necessary to use microcombs with low phase noise, high coherence, and stable mode locking. A number of mode-locking approaches have been reported [17, 18]. It is worth noting that even with relatively incoherent microcombs, processors can still achieve an acceptable accuracy because the microcomb mainly serves as a multi-wavelength source and the optical powers of different wavelength channels are detected incoherently by a BPD.

B. *Influence of the electro-optic modulator*

In Fig. 2, an electro-optic modulator is used to modulate the input microwave signal onto different wavelength channels. The most commonly used electro-optic modulators are Mach-Zehnder modulators (MZMs), owing to their high modulation efficiency, low insertion loss, and large operation bandwidth [46]. Due to the asymmetry in the electric field overlap at each electrode [47], practical MZMs not only produce intensity modulation, but also give rise to undesired phase modulation, known as modulation chirp. The chirp leads to distortions in the modulated optical signals, thus resulting in processing errors. Here, we analyze the influence of modulator chirp on the accuracy for different processing functions.

The chirp of a MZM can be characterized by the chirp parameter given by [48]

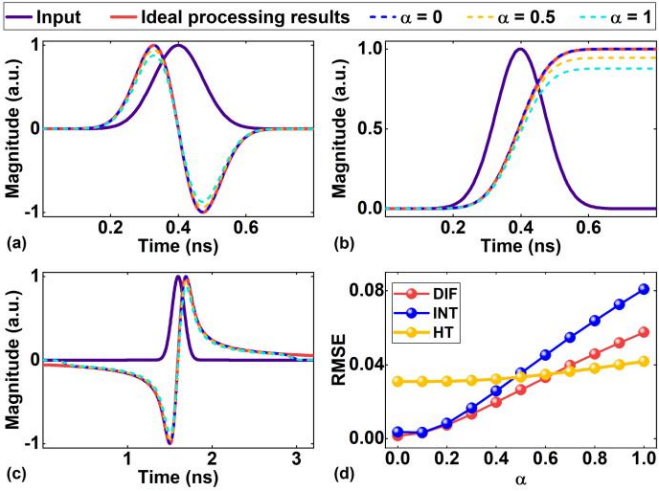

Fig. 5. Influence of the modulator chirp on errors of differentiation (DIF), integration (INT), and Hilbert transformation (HT). (a) – (c) Temporal waveform of Gaussian input pulse and output waveforms from the transversal signal processors performing (a) DIF, (b) INT, and (c) HT. Different curves show the results for different chirp parameter $\alpha$. The ideal processing results are also shown for comparison. (d) Corresponding RMSEs between the ideal results and the processors' output waveforms as a function of $\alpha$. In (a) – (d), the Gaussian input pulse has a FWHM of ~0.17 ns. The tap number, comb spacing, length of dispersive medium, and SOD parameter are $M$ = 80, $\Delta\lambda$ = 0.4 nm, $L$ = 4.8 km, and $D_2$ = 17.4 ps/nm/km, respectively.

$$\alpha = \frac{\gamma_1 + \gamma_2}{\gamma_1 - \gamma_2} \qquad (8)$$

where $\gamma_1$ and $\gamma_2$ are the voltage-to-phase conversion coefficients for the two arms of the MZM. When $\alpha$ = 0 (*i.e.*, $\gamma_1$ = $-\gamma_2$), pure intensity modulation is achieved. Figs. 5(a) – (c) show the output waveforms from microcomb-based MWP transversal signal processors that perform DIF, INT, and HT for different chirp parameters $\alpha$. The ideal processing result without theoretical errors and the results that only account for theoretical errors (corresponding to $\alpha$ = 0) are also shown for comparison. For all processing functions the output waveforms approach the ideal results as $\alpha$ decreases from 1 to 0, indicating the reduced system error for a lower modulator chirp.

Fig. 5(d) shows the calculated RMSEs versus modulator chirp $\alpha$. As expected, the RMSE increases with $\alpha$ for all processing functions, which agrees with the trend in Figs. 5(a) – (c). We also noted that the impact of the modulation chirp on the system performance is more significant for the DIF and INT functions as compared to the HT.

### C. Influence of the single-mode fibre

In Fig. 2, a spool of SMF is employed as the dispersive module of the MWP transversal signal processor, which introduces both amplitude and phase errors due to its chromatic dispersion, including both SOD and TOD. SOD induces a uniform time delay between adjacent taps, which is required for MWP transversal signal processors without alignment errors. However, SOD also introduces a time delay between the modulated sidebands, which leads to a power degradation of the microwave output after photodetection, and hence system errors [49]. On the other hand, the SMF TOD introduces non-uniform time delays between adjacent taps, thus resulting in undesired phase errors. In this section, we analyze the influence of the SMF's SOD and TOD on the accuracy for different processing functions.

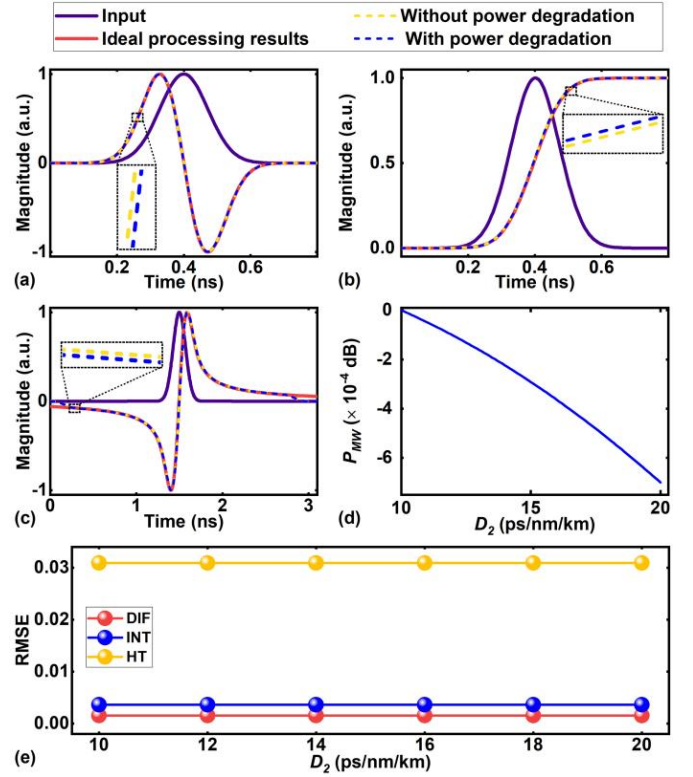

Fig. 6. Influence of SMF's SOD on errors of differentiation (DIF), integration (INT), and Hilbert transformation (HT). (a) – (c) Temporal waveform of Gaussian input pulse and output waveforms from the transversal signal processors performing (a) DIF, (b) INT, and (c) HT. Different curves show the results with and without the influence of power degradation induced by SOD. The SOD parameter is $D_2$ = 17.4 ps/nm/km. The ideal processing results are also shown for comparison. (d) Power degradation of the output microwave signal $P_{MW}$ as a function of the SOD parameter $D_2$. (e) Corresponding RMSEs between the ideal results and the processors' output waveforms as a function of $D_2$. In (a) – (e), the Gaussian input pulse has a FWHM of ~0.17 ns. The tap number, comb spacing, and length of dispersive medium are $M$ = 80, $\Delta\lambda$ = 0.4 nm, and $L$ = 4.8 km, respectively.

A MZM generates two modulated sidebands, with the output termed a double-sideband (DSB) signal. The SOD of the SMF generates different phase shifts for the two sidebands resulting in different phase shifts between the carrier and the two beat microwave sidebands. Therefore, the final microwave output after photodetection experiences a power degradation, with its power given approximately by [49]

$$P_{MW} \propto \cos\left(\frac{\pi L D_2}{c} \lambda_c^2 f_{MW}^2\right) \qquad (9)$$

where $c$ is the speed of light in vacuum, $\lambda_c$ is the center wavelength of each channel, and $f_{MW}$ is the frequency of the input microwave signal.

Figs. 6(a) – (c) show the output waveforms from the processors for the DIF, INT, and HT functions, with and without including the power degradation caused by SOD. The SOD parameter is kept constant at $D_2$ = 17.4 ps/nm/km. For all processing functions, there are only slight differences induced

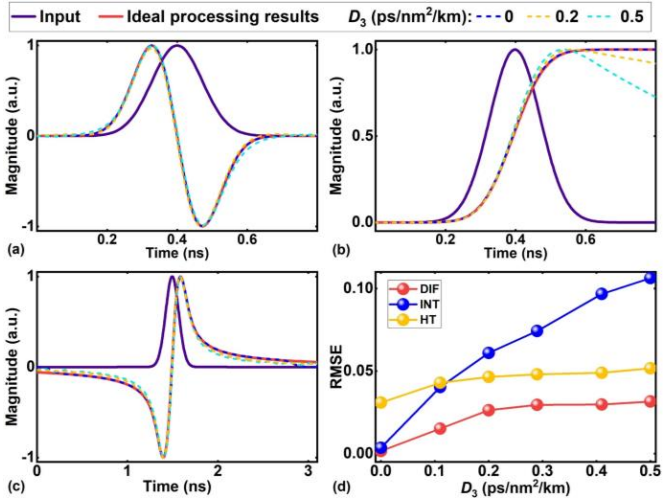

Fig. 7. Influence of SMF's TOD on errors of differentiation (DIF), integration (INT), and Hilbert transformation (HT). (a) – (c) Temporal waveform of Gaussian input pulse and output waveforms from the transversal signal processors performing (a) DIF, (b) INT, and (c) HT. Different curves show the results for different TOD parameter $D_3$. The ideal processing results are also shown for comparison. (d) Corresponding RMSEs between the ideal results and the processors' output waveforms as a function of $D_3$. In (a) – (d), the Gaussian input pulse has a FWHM of ~0.17 ns. The tap number, comb spacing, length of dispersive medium, and SOD parameter are $M = 80$, $\Delta\lambda = 0.4$ nm, $L = 4.8$ km, and $D_2 = 17.4$ ps/nm/km, respectively.

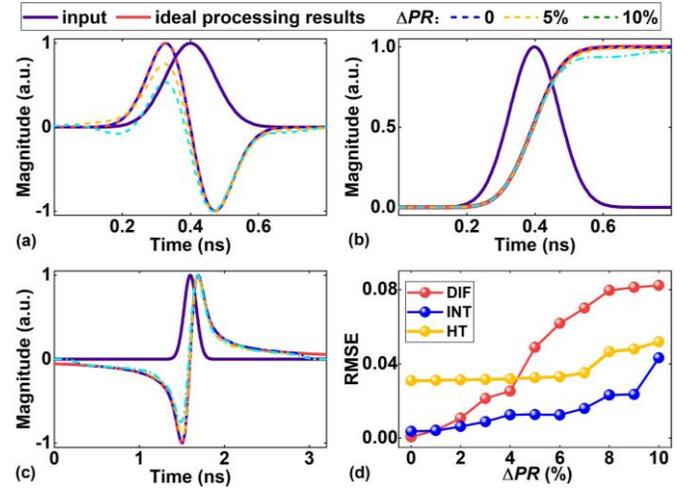

Fig. 8. Influence of shaping errors induced by the OSS on accuracy of differentiation (DIF), integration (INT), and Hilbert transformation (HT). (a) – (c) Temporal waveform of Gaussian input pulse and output waveforms from the transversal signal processors performing (a) DIF, (b) INT, and (c) HT. Different curves show the results for different percentage ranges ($\Delta PRs$) of random tap coefficient errors (RTCEs). The ideal processing results are also shown for comparison. (d) Corresponding RMSEs between the ideal results and the processors' output waveforms as a function of $\Delta PR$. In (a) – (d), the Gaussian input pulse has a FWHM of ~0.17 ns. The tap number, comb spacing, length of dispersive medium, and SOD parameter are $M = 80$, $\Delta\lambda = 0.4$ nm, $L = 4.8$ km, and $D_2 = 17.4$ ps/nm/km, respectively.

by SOD. Fig. 6(d) shows the power degradation $P_{MW}$ as a function of $D_2$, which is calculated based on Eq. (9). As can be seen, the power degradation induced by SOD is very small, being < $10^{-3}$ dB for $D_2 = 17.4$ ps/nm/km in Figs. 6(a) – (c).

Fig. 6(e) shows the RMSE as a function of $D_2$, showing that the RMSEs only vary very slightly (< $10^{-4}$) with $D_2$ for all processing functions, in agreement with Figs. 6(a) – (c). These results indicate that although the SOD of SMF induces power degradation of the microwave output, its influence on the system accuracy is very small.

The TOD of the SMF introduces additional non-uniform time delays between the modulated replicas in the wavelength channels, thus resulting in alignment errors in the processing results. The additional time delay of the $n^{th}$ tap is given by [50]

$$\Delta T_{TOD} = D_3 \, L \, \Delta\lambda^2 \, n^2 \quad (10)$$

where $D_3$ is the TOD parameter.

Figs. 7(a) – (c) show the output waveforms from processors that perform DIF, INT, and HT, versus the TOD parameter $D_3$. The ideal processing result without theoretical errors and the results that only account for theoretical errors (corresponding to $D_3 = 0$) are also shown for comparison. For all processing functions, the processors' outputs approach the ideal processing results as $D_3$ decreases from 0.5 ps/nm²/km to zero, indicating that improved accuracy can be achieved for a smaller TOD.

Fig. 7(d) shows the RMSE as a function of $D_3$, where, as expected, the RMSE increases with increasing $D_3$ for all functions – agreeing with the trend in Figs. 7(a) – (c). The influence of TOD on the system performance is more significant than that of the SOD. We also note that the INT function is more susceptible to errors induced by the TOD as compared to the DIF and HT functions, reflecting that INT has a more stringent requirement for the accuracy of the phase of the different taps.

### D. Influence of optical spectral shapers and photodetectors

In Fig. 2, an OSS is used as a spectral shaping module to weight the delayed signals across different wavelength channels according to the designed tap coefficients. This is followed by a BPD that sums the delayed and weighted signals to generate the microwave output of the processor. The OSS induces shaping errors, which result in inaccurate tap coefficients and hence output errors. On the other hand, noise and an uneven transmission response of the BPD lead to variations of the power of the microwave output. In this section, we analyze the influence of these error sources for the different processing functions.

We introduce random tap coefficient errors (RTCEs) within a certain percentage range of $\Delta PR$ to characterize the shaping errors of the OSS. Figs. 8(a) – (c) show the output waveforms from the processors for all functions and for the RTCEs in different ranges, together with the ideal processing result without theoretical errors and the results that only account for theoretical errors (corresponding to $\Delta PR = 0$). For all the three processing functions, the processors' output waveforms show better agreement with the ideal results for a smaller $\Delta PR$, reflecting an improved accuracy associated with reduced RTCEs.

Fig. 8(d) shows the RMSE as a function of $\Delta PR$, showing that the RMSE increases with $\Delta PR$ for all functions, agreeing with the trend in Figs. 8(a) – (c). The shaping errors of the OSS have a more obvious impact on the accuracy for DIF as compared to the other two functions, indicating that DIF has a more stringent requirement for the accuracy of the tap amplitudes.

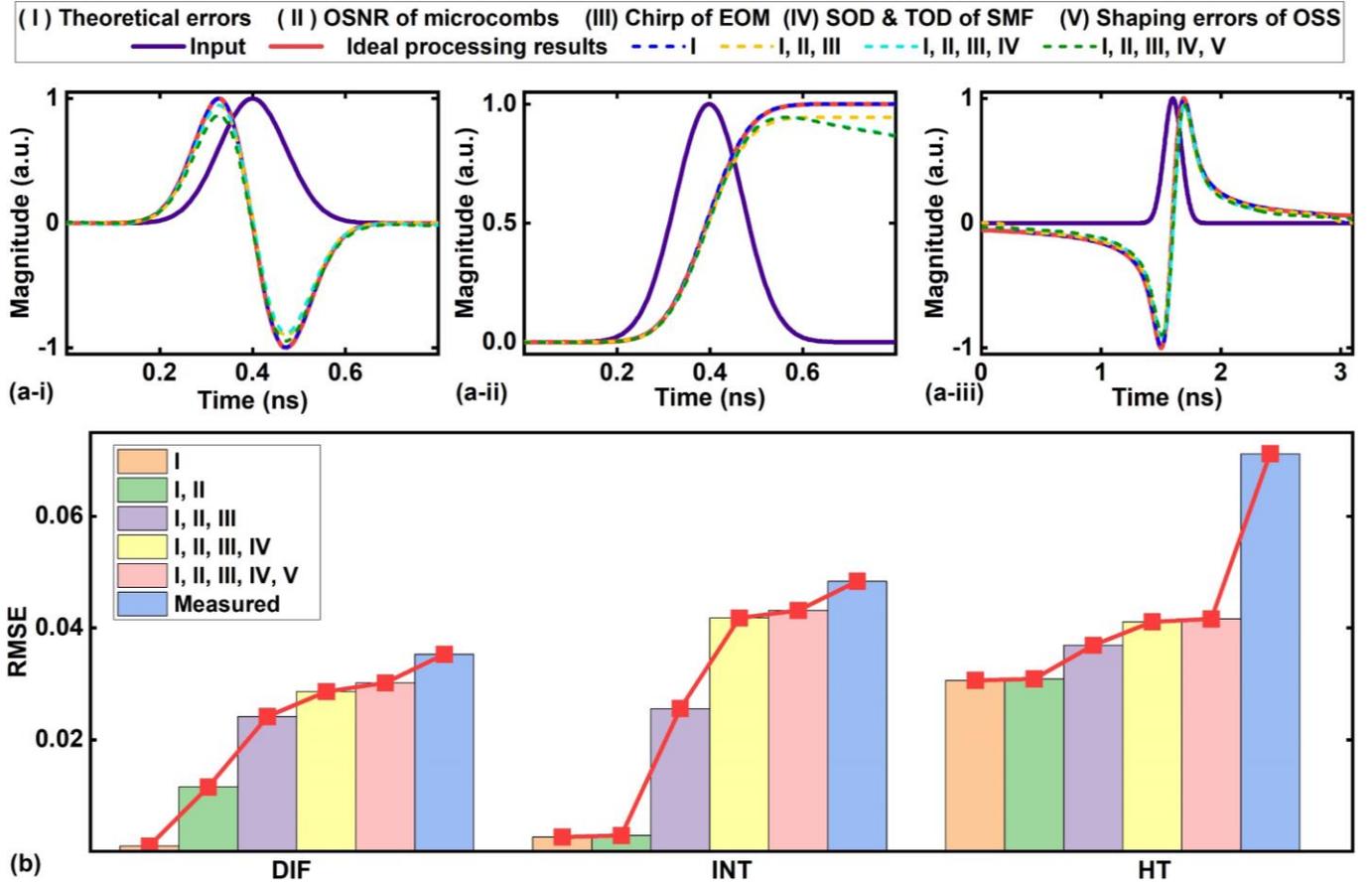

Fig. 9. Contributions of different error sources to the overall errors of differentiation (DIF), integration (INT), and Hilbert transformation (HT). (a) Temporal waveform of Gaussian input pulse and output waveforms from the transversal signal processors performing (i) DIF, (ii) INT, and (iii) HT. Different curves show the results after accumulating errors induced by different sources from I to V. The ideal processing results are also shown for comparison. (b) Corresponding RMSEs between the ideal results and the processors' outputs. The practical measured RMSEs are also shown. In (a) and (b), the microcomb has an OSNR of 30 dB. The chirp parameter, SOD parameter, TOD parameter, and tap coefficient fluctuations are $\alpha = 0.5$, $D_2 = 17.4$ ps/nm/km, $D_3 = 0.083$ ps/nm$^2$/km, and $\Delta PR = 5\%$. The Gaussian input pulse has a FWHM of ~0.17 ns. The tap number, comb spacing, and length of dispersive medium are $M = 80$, $\Delta\lambda = 0.4$ nm, and $L = 4.8$ km respectively.

In Fig. 2, the use of a BPD greatly suppresses the common-mode noise of the optical signal, which largely cancels out the intensity noise caused by the photodetector. Therefore, the errors induced by the BPD mainly come from its limited response bandwidth and uneven transmission response, which introduce additional errors in the tap coefficients after spectral shaping. Similarly, the limited bandwidth and uneven response of the EOM could also introduce additional errors to the tap coefficients before spectral shaping. These errors, together with the shaping errors of the OSS, can be effectively mitigated through feedback control, which will be discussed in section IV. Finally, we note that the BPD shot noise can induce random power fluctuations in the output microwave signal which limits the lowest achievable phase noise floor [51]. The influence of this on the system performance is similar to the microcomb noise, and can be reduced by using a BPD with higher sensitivity [52].

*E. Contributions of different error sources*

In this section, we analyze the contribution of the error sources discussed above to the overall processing errors of microcomb-based MWP transversal signal processors and provide a global picture to show the impact of different error sources.

Fig. 9(a) shows the simulated output waveforms for all functions, including errors induced by the sources from I to V in Fig. 2, with the ideal results shown for comparison. Based on the measurements and parameters of the components in our previous experiments [27, 30, 34], the chirp parameter of the EOM, SOD and TOD parameter of the SMF, and range of RTCEs are set to $\alpha = 0.5$, $D_2 = 17.4$ ps/nm/km, $D_3 = 0.083$ ps/nm$^2$/km, and $\Delta PR = 5\%$, respectively. In our simulations, we also used the OSNRs of the comb lines that were measured by an optical spectrum analyzer. As expected, the overall output errors become larger with the accumulation of errors induced by these sources for all processing functions.

In order to quantify the contributions of the different sources of error, we calculate the RMSEs from the simulation results Fig. 9(a) and plot them in Fig. 9(b). The experimentally measured RMSEs are also shown for comparison. In our simulations, we used the input microwave signal waveform measured by a high-bandwidth real-time oscilloscope to calculate the RMSEs, this can minimize the errors induced by the discrepancy between the experimentally generated and ideal Gaussian pulses. The RMSEs of the simulation results increase with the accumulation of errors, which agrees with the trend in Fig. 9(a). There are margins between the RMSEs of the simulation results and the experimental results. They are mainly caused by deviations between the simulation and experiment parameters as well as factors that are not accounted for in our simulation, such as the phase noise of the microcomb, the limited response bandwidth and uneven transmission response of the EOM and BPD, and the shot noise of the BPD. As shown in Fig. 9(b), different processing functions show distinct errors induced by the experimental imperfections. This is mainly induced by the differences in their spectral transfer functions, as indicated by Eqs. (5) – (7), which lead to different responses to the experimental error sources. As can be seen, the system error for the DIF is mainly induced by the microcomb imperfections and EOM chirp. For the INT, the main error sources are the EOM chirp and the SMF TOD. As compared to the DIF and INT, the theoretical errors have a more significant influence on the accuracy for the HT.

TABLE I. COMPARISON OF COMPONENTS' PARAMETERS IN DISCRETE AND INTEGRATED PROCESSORS

| | No. | Tap No. | OSNR of microcombs | Chirp parameter of the EOM | Errors of the delay element | RTCE of the spectral shaping module |
|---|---|---|---|---|---|---|
| Discrete processor | 1 | $M = 80$ [41] | $OSNR$ : 20 dB [41] | $\alpha$ : 0.1 [53] | $t_v$ : 4% [41] | $RTCE$ : 5% [41] |
| Integrated processors | No. | Tap No. | OSNR of microcombs | Chirp parameter of the EOM | Error of the delay element | RTCE of the spectral shaping module |
| | 2 | $M = 8$ [54] | $OSNR$ : 20 dB [41] | $\alpha$ : 0.8 [48] | $t_v$ : 3% [55] | $RTCE$ : 9% [56] |
| | No. | Tap No. | OSNR of microcombs | Chirp parameter of the EOM | Error of the delay element | RTCE of the spectral shaping module |
| | 3 | $M = 20$ | $OSNR$ : 20 dB [41] | $\alpha$ : 0.8 [48] | $t_v$ : 3% [55] | $RTCE$ : 9% [56] |

*F. Performance comparison of processors implemented by discrete versus integrated components*

Early implementations of microcomb-based MWP transversal signal processors simply replaced conventional multi-wavelength sources with optical microcombs while retaining all other components as discrete devices [18, 36]. Recently, several processors comprised entirely of integrated components have also been demonstrated [54, 55]. Despite being based on the same operation principle, the processors implemented with discrete versus integrated components exhibit different processing performance. In this section, we compare their processing accuracy. To simplify our discussion, we refer to the processors implemented in these two forms as discrete versus integrated processors.

Table I summarizes parameters of the components in three processors that we investigate, including a discrete processor (Processor 1) and two integrated processors (Processors 2 and 3). There are two integrated processors: one with the same tap number as that in Ref. [54] and the other with an increased tap number to demonstrate the potential for improvement. Although the size, weight, and power consumption (SWaP) of integrated processors are greatly reduced compared with the discrete processors, the state-of-the-art integrated processors suffer from limited tap numbers due to the restrictions imposed by the integrated components. Currently, integrated processors with only 8 [54] and 12 taps [55] have been demonstrated, whereas discrete processors have been implemented with up to 80 taps [18, 41]. To characterize the errors induced by imperfect response of experimental components, OSNR of microcombs, chirp parameters ($\alpha$), error of the delay element ($t_v$), and random tap coefficient errors (RTCEs) induced by the spectral shaping module were introduced. All of these parameters were set based on the practical processors in Refs. [41, 48, 53, 55, 56]. For comparison, we assumed that the three processors have the same comb spacing of ~0.4 nm and the same time delay between adjacent taps of $\Delta T$ = ~33.4 ps.

Figs. 10(a) – (c) show the outputs of Processors 1 – 3 in Table I that perform DIF, INT, and HT, respectively. Here we show the processors' outputs with errors induced by (1) only limited tap numbers and (2) both limited tap numbers and experimental errors. The ideal processing results are also shown for comparison. Deviations between the processors' outputs and the ideal results are observed for all three functions, and the deviations become more significant when taking into account the experimental errors. Fig. 10(d) compares the RMSEs of the processors in Figs. 10(a) – (c). The higher processing accuracy of the discrete processor, compared to the integrated processors, is reflected by the lower RMSEs of Processor 1 for all three processing functions. In addition, the RMSEs of Processor 3 are lower compared to Processor 2, which indicates a higher processing accuracy achieved by increasing the tap number. According to Fig. 10(d), the primary factor that contributes to the degradation of accuracy for integrated processors is the limited tap number. Whereas for discrete processors with a sufficiently large tap number, the processing inaccuracy is mainly induced by the imperfect response of experimental components. We also note that the differences in RMSEs among Processors 1 – 3 are more prominent for the INT than

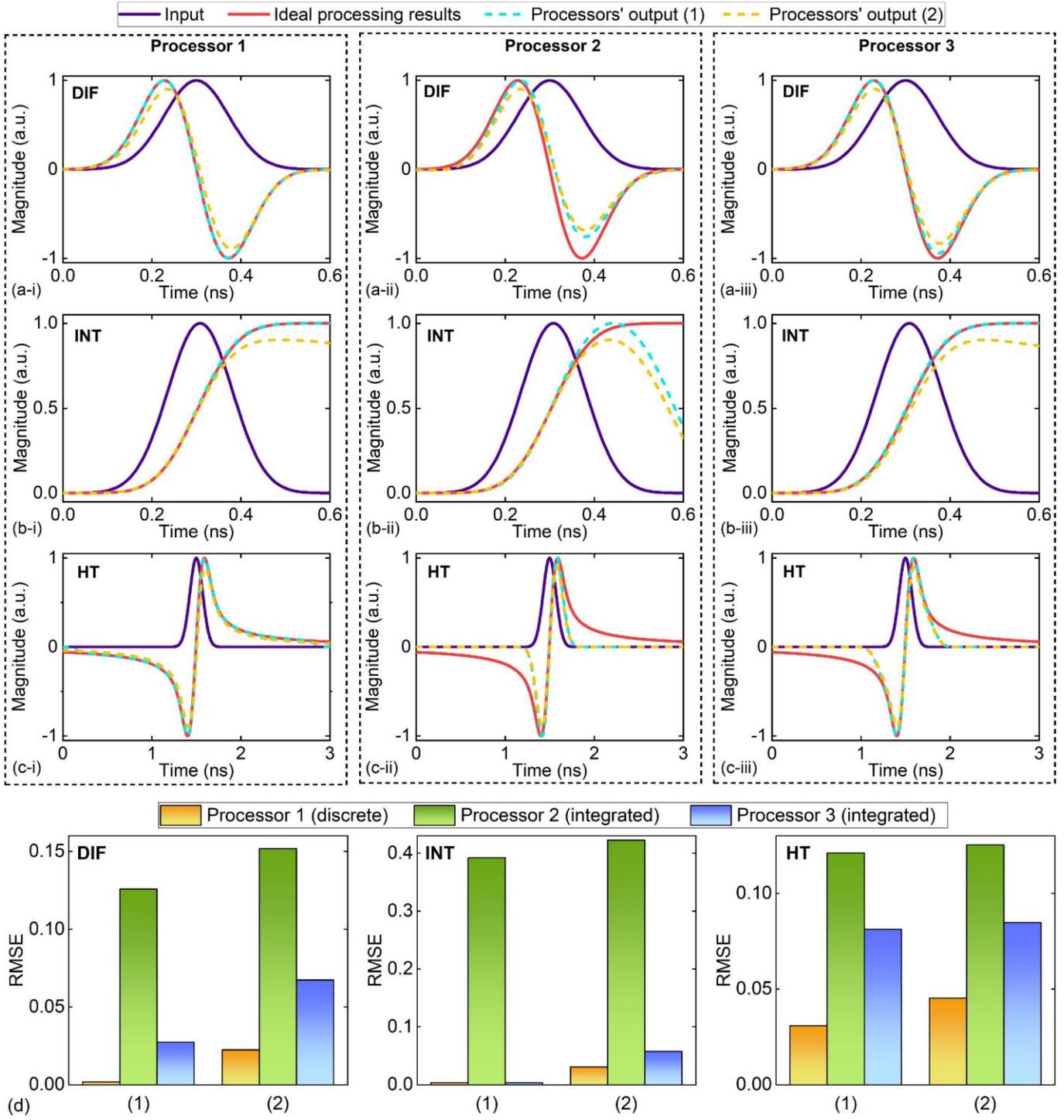

Fig. 10. Temporal waveform of Gaussian input pulse and output waveforms from Processors 1 – 3 that perform (a) differentiation (DIF), (b) integration (INT), and (c) Hilbert transform (HT). (d) Comparison of corresponding RMSEs. In (a) – (d), we show the results with errors induced by (1) only limited tap numbers and (2) both limited tap numbers and experimental errors, together with the ideal processing results for comparison.

the other two processing functions, indicating a higher requirement for a greater number of taps to improve the processing accuracy of INT. In addition, experimental errors have a substantial impact on the RMSEs of DIF, whereas their impact on HT is very small.

## IV. ERROR COMPENSATION VIA FEEDBACK CONTROL

In this section, feedback control is introduced to compensate for errors induced by the imperfect response of experimental components. The benefit of feedback control is quantitatively analyzed by comparing the system errors with and without feedback control.

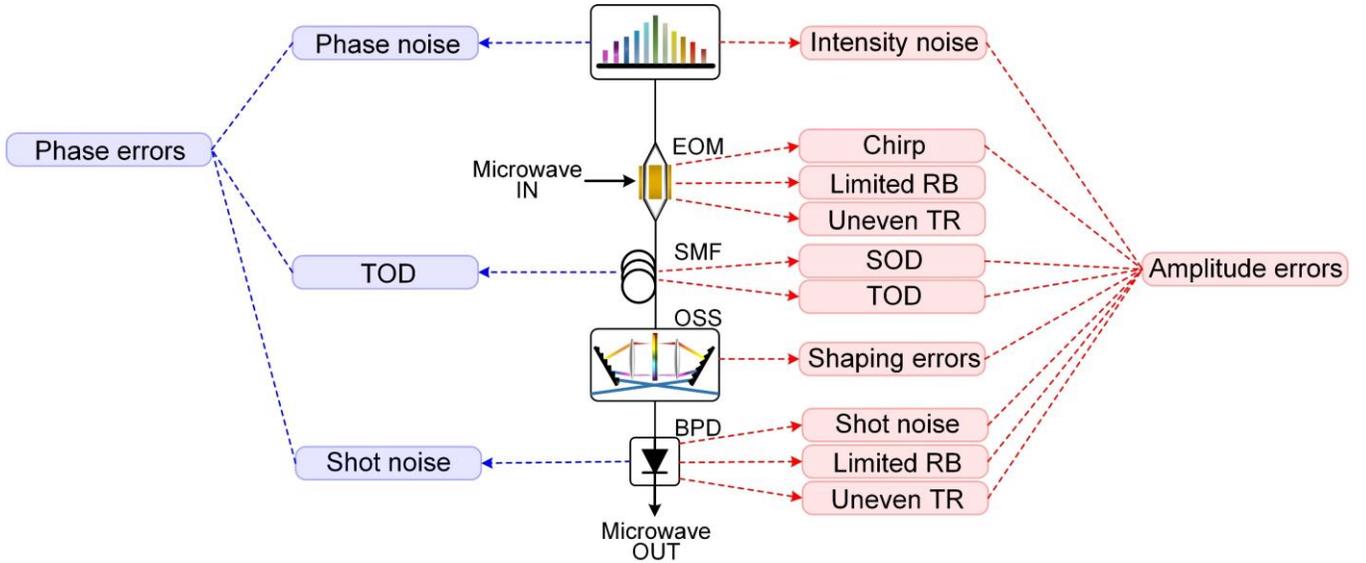

Fig. 11. Amplitude and phase errors induced by different components in microcomb-based MWP transversal signal processors. EOM: electro-optic modulator. SMF: single-mode fiber. OSS: optical spectral shaper. BPD: balanced photodetector. RB: response bandwidth. TR: transmission response. SOD: second-order dispersion. TOD: third-order dispersion.

As shown in Fig. 11, we classify the error sources discussed in Section III into two categories, depending on whether amplitude or phase errors are introduced in the taps. The amplitude and phase errors refer to errors in the tap coefficients (i.e., $a_n$ in Eqs. (1) – (3)) and time delays (i.e., $n\Delta T$ in Eqs. (1) – (3)) for different taps, respectively. The sources of amplitude errors include the microcomb intensity noise, EOM chirp, TOD and SOD of the SMF, OSS shaping errors, BPD shot noise, and the bandwidth response of the EOM and BPD. The sources of phase errors include microcomb phase noise, TOD of the SMF, and BPD shot noise. We note that some of the error sources in Fig. 11 are static or slowly varying, e.g., chirp of EOM, SOD and TOD of SMF, and shaping errors of OSS. In contrast, the fluctuations in the amplitude and phase caused by microcombs and the BPD are normally faster – on the order of 10 GHz.

The static and slowly varying errors in Fig. 11 induced by different error sources can be compensated for by introducing feedback control to calibrate the tap coefficients set for the OSS. Fig. 12(a) shows a schematic of a MWP transversal signal processor with feedback control. A feedback control loop including all the components of the signal processor is introduced to calibrate both the amplitude and phase of each comb line based on the ideal impulse response. This allows for the compensation of the errors induced by different components in the feedback loop. During the amplitude calibration process, a microwave signal is employed as the input signal to test the impulse response of the processor channel by channel, where the same input microwave signal is modulated onto the corresponding comb line. The intensities of the microwave signals after photodetection are recorded by an oscilloscope and sent to a computer, where they are subtracted from the designed tap weights to generate error signals. Finally, the generated error signals are sent to the OSS to calibrate the attenuation of comb line intensity. After several iterations of the above process, the amplitude errors caused by the non-ideal impulse response of the system can be effectively reduced. Similarly, the static and slowly varying phase errors can be mitigated by exploiting the programmable phase characteristics of the OSS to compensate the deviation between the measured and desired phase response.

In Fig. 12(b), we compare the RMSEs for all functions with and without feedback control. The RMSEs caused by theoretical errors are also shown for comparison. As expected, the measured RMSEs with feedback control are much lower than those measured without calibration and approach the theoretical RMSEs more closely. After calibration, there are still discrepancies between the measured RMSEs and theoretical RMSEs, reflecting that there are still residual errors that cannot be compensated for with feedback control. We infer that these errors are mainly induced by rapidly varying error sources, by deviations between the simulated and experimental parameters, and by the limited resolution of the instruments such as the OSS and oscilloscope.

To further improve the system accuracy, multiple-stage feedback control can be employed. For example, another feedback loop with one more OSS can be introduced in the microcomb generation module to flatten the comb lines of the initially generated microcomb. This allows for uniform wavelength channel link gain and can also reduce the loss control range for the spectral shaping in the transversal signal processing module. Recently, self-calibrating photonic integrated circuits have been demonstrated [57, 58], where the impulse response calibration was achieved by incorporating an optical reference path to establish a Kramers-Kronig relationship and then calculate the amplitude and phase errors based on a Fourier transform. This offers new possibilities to achieve precise feedback control in microcomb-based MWP transversal signal processors.

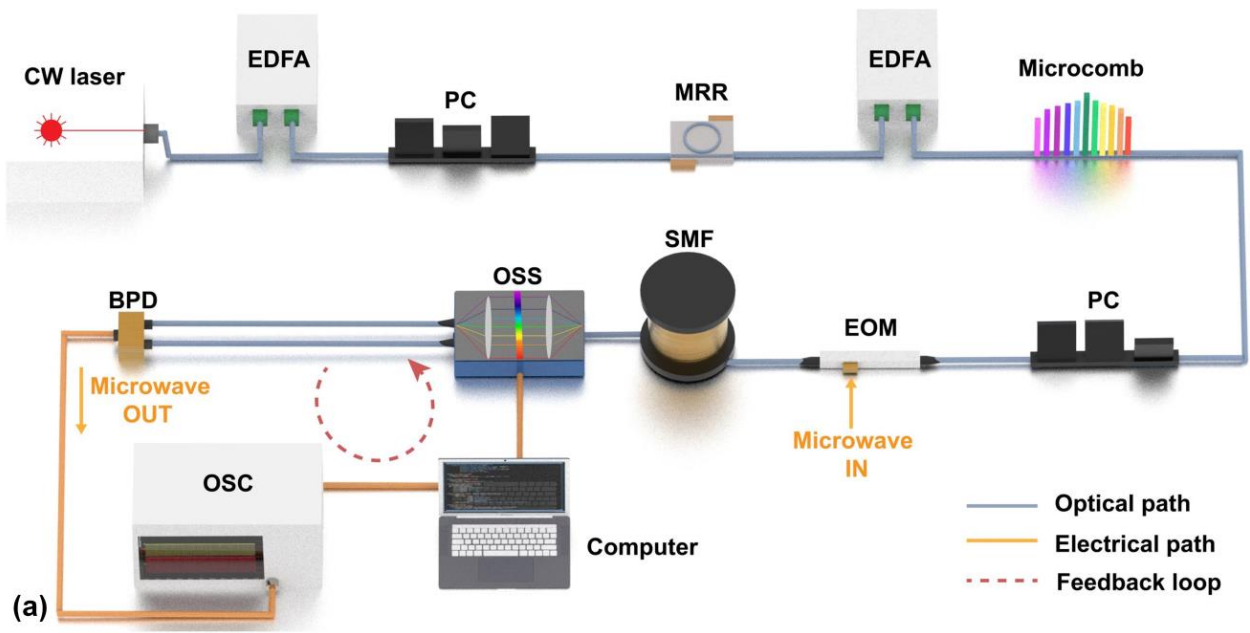

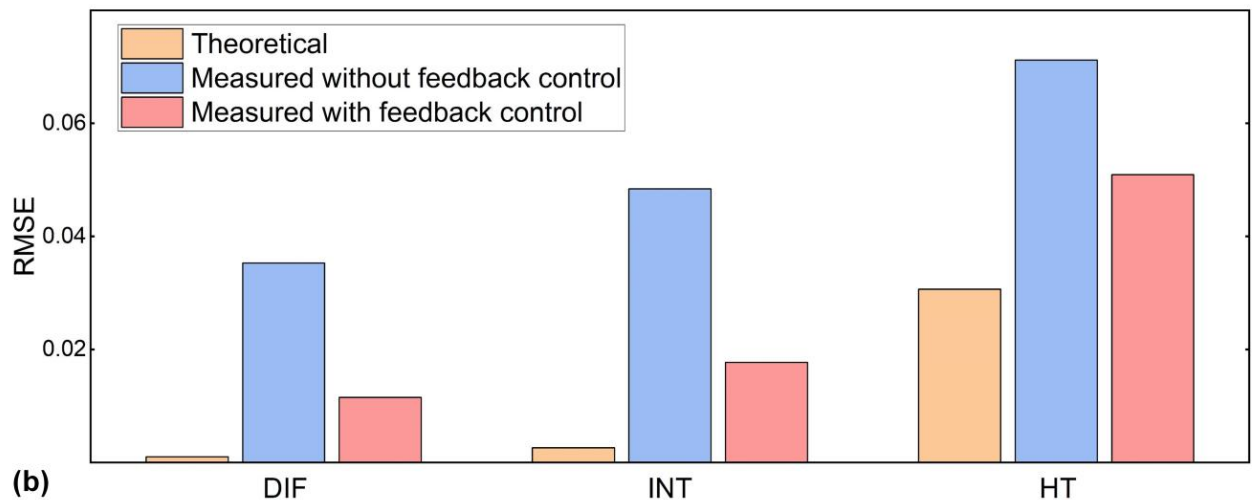

Fig. 12. (a) Schematic of a microcomb-based MWP transversal signal processor with feedback control. CW laser: continuous-wave laser. EDFA: erbium-doped fiber amplifier. PC: polarization controller. MRR: microring resonator. OSS: optical spectral shaper. OC: optical coupler. EOM: electro-optic modulator. SMF: single-mode fiber. BPD: balanced photodetector. OSC: oscilloscope. (b) Comparison of measured RMSEs for DIF, INT, and HT with and without feedback control. The corresponding theoretical RMSEs are also shown for comparison. The tap number, comb spacing, length of dispersive medium, and SOD parameter are $M = 80$, $\Delta\lambda = 0.4$ nm, $L = 4.8$ km, and $D_2 = 17.4$ ps/nm/km, respectively. The input microwave signals are Gaussian pulses with a FWHM of ~0.17 ns.

Apart from implementing feedback control, there are some other methods to reduce the errors induced by experimental imperfections. For example, employing advanced mode-locking approaches [18] to reduce the noise of microcombs could be beneficial for both discrete and integrated processors. For integrated processors, the chirp of silicon EOM can be mitigated by using push-pull configurations as well as p-n depletion mode structure [53], and proper methods to calibrate the bias point [55]. The shaping errors of integrated spectral shapers can be alleviated via calibration procedures and gradient-descent control [55]. Integrated delay elements introduce additional loss especially when using a waveguide with high propagation loss, and adiabatic Euler bends can be employed to achieve low-loss and low-crosstalk waveguide bends [59]. The use of a wavelength-addressable serial integration scheme can also enable large-scale integration [60].

## V.  CONCLUSION

In summary, we analyze the processing errors induced by experimental imperfections for microcomb-based MWP transversal signal processors. We first investigate the errors arising from imperfect microcomb characteristics, EOM chirp, chromatic dispersion in the dispersive module, errors in the OSS, and photodetector noise. Next, we present a global picture of the quantitative influence of different error sources on the overall system performance. Finally, we introduce feedback control to compensate for the errors and quantitatively analyze the improvement in the processing accuracy. Our results show that the influence of the error sources varies for the different processing functions studied here, and that these errors can be significantly reduced by introducing feedback control for both static and slowly varying sources of error. This work provides a useful guide for optimizing the performance of microcomb-based MWP transversal signal processors for versatile high-speed information processing applications.

**Yang Sun** received her bachelor's degree from the Beijing Institute of Technology, China in June 2016 and her master's degree from the Beijing University of Posts and Telecommunications, China in June 2019. She is currently a Ph.D. candidate at the Optical Sciences Centre in Swinburne University of Technology under the supervision of Prof. David J. Moss and Dr. Jiayang Wu. Her current research focuses on integrated nonlinear optics, microwave photonics, and neuromorphic computing.

**Jiayang Wu** (S'12–M'17–SM'22) received the B.Eng. degree in communication engineering in September 2010 from Xidian University, Xi'an, China, and the Ph.D. degree in electronics engineering in December 2015 from Shanghai Jiao Tong University, Shanghai, China. After that, he joined the Swinburne University of Technology and became a Postdoctoral Research Fellow in 2016. He is currently a Senior Research Fellow and a Senior Lecturer at the Optical Sciences Centre, Swinburne University of Technology. His current research fields include integrated photonics, nonlinear optics, 2D materials, and optical communications. As of August 1st of 2023, he has over 220 refereed publications in SCI journals and EI conference proceedings, highlighted by *Nature*, *Nature Reviews Chemistry*, *Advanced Materials*, *Advances in Optics and Photonics*, *Nature Communications*, *Applied Physics Reviews*, *Small*, *Small Methods*, *Laser & Photonics Reviews*, and *Nano Letters*. He is also an inventor on 15 filed patents about integrated photonic devices and optical communication technologies. In 2021, Dr. Wu was awarded the Australian National Research Award as the top researcher in the field of Optics & Photonics (only 1 person). In 2021 and 2022, he also ranked in the top 2% global highly cited researchers in the discipline of Optics (Elsevier science-wide author databases of standardized citation indicators).

**Yang Li** received his bachelor's degree from the University of Jinan, China in June 2015 and his master's degree from the University of New South Wales, Australia in January 2021. He is currently a Ph.D. candidate under the supervision of Prof. David J. Moss and Dr. Jiayang Wu, in the Optical Sciences Centre, Swinburne University of Technology. His current research focuses on integrated nonlinear optics, microwave photonics, and neuromorphic computing.

**Xingyuan Xu** is currently a professor with Beijing University of Posts and Telecommunications. He received his Ph.D. from Swinburne University of Technology, supervised by Prof. David J. Moss. His research focuses on neuromorphic optics, optical signal processing, and optical frequency combs. He was awarded 2019 IEEE Photonics Society Graduate Student Scholarship, 2020 Iain Wallace Research Medal from Swinburne University of Technology, 2020 Extraordinary Potential Prize of Chinese Government Award for Outstanding Self-financed Students Abroad, and 2021 *IEEE Journal of Lightwave Technology* best paper award.

**Guanghui Ren** received the Ph.D. degree in 2016 from RMIT University, Melbourne, VIC, Australia, where he is currently working as a Senior Research Fellow in the Integrated Photonics and Applications Centre (InPAC). His research interests include integrated optics, silicon photonics, thin-film lithium niobate, bio-photonics and hybrid integration of functional materials on integrated optics platform.

**Mengxi (Sim) Tan** (S'17) has been a Research Fellowat RMIT University since 2021. She received the B.Eng. degree from Changchun University of Science and Technology in opto-electronic information engineering in 2014, a M.S. degree in optical engineering in 2017 from Beijing University of Aeronautics and Astronautics, and a Ph.D. degree at the Swinburne University of Technology, Melbourne, VIC, Australia, in 2021. In 2021 she received an IEEE Photonics society graduate student scholarship: 1 of only 10 awarded worldwide each year. She won the 2021 Swinburne HDR Woman Research of the Year. Her current research interests include integrated nonlinear optics and high-speed optical signal processing, microwave photonics, and optical neuromorphic processing. She has published papers in *Nature* (2021) and *Nature Communications* (2020). She is a member of the IEEE Photonics Society and Optica.

**Sai Tak Chu** received the B,Sc. Degree in computing and computer electronics from Wilfrid Laurier University, Canada and the M.Sc. degree in physics and Ph.D. degree in electrical engineering from the University of Waterloo, Canada, in 1984, 1986 and 1990, respectively. He has been involved in the research, development and commercialization of waveguide based photonics devices for over twenty-five years. Prior to joining the Department of Physics, City University of Hong Kong, Hong Kong, as an Associate Professor in 2010, he spent the 1990's carried out research in a number of research institutes, including CITRC Canada, KAST, Japan and NIST, USA. In 2000 he co-founded Little Optics Inc. in the USA. An optical component company specialized in densely integrated PLC for broadband communication. The company was subsequently acquired by Infinera, USA in 2006. His research interests include the areas of integrated photonic devices, biophotonics and terahertz technologies.

**Brent E. Little** biography is not available at the time of submission.

**Roberto Morandotti** (M'09–SM'14–F'21) received the M.Sc. degree in physics from the University of Genova, Genova, Italy, in 1993, and the Ph.D. degree from the University of Glasgow, Glasgow, UK, in 1999. From 1999 to 2001, he was with the Weizmann Institute of Science, Rehovot, Israel. From 2002 to 2003, he was with the University of Toronto, ON, Canada, where he was involved in the characterization of novel integrated optical structures. In June 2003, he joined the Institut National de la Recherche Scientifique, Centre Énergie Matériaux Télécommunications, Varennes, QC, Canada, where he has been a Full Professor since 2008. He is the author and coauthor of more than 700 papers published in international scientific journals and conference proceedings. His current research interests include the linear and nonlinear properties of various structures for integrated and quantum optics, as well as nonlinear optics at unusual geometries and wavelengths, including terahertz. He is a Fellow of IEEE, of the Royal Society of Canada, of the American Physical Society, of the Optica (formerly the OSA), of the SPIE, and an E.W.R Steacie Memorial Fellow. He served as a Chair and Technical Committee Member for several Optica, IEEE, and SPIE sponsored meetings.

**Arnan Mitchell** was awarded the Ph.D. in engineering in 2000 from RMIT University. He is the leader of the Microplatforms research group and Director of the ARC Centre of Excellence for optical microcombs for breakthrough science. He was Chief Investigator and RMIT Node Director of the ARC Centre of Excellence for ultrahigh bandwidth devices for optical systems (CUDOS). His research is highly multidisciplinary, spanning microfluidics, integrated optics, photonic signal processing, functional materials, microsystems, nanomaterials, and lab-on-a-chip technology. He and his team focus on platforms that enable fundamental scientific and biomedical breakthroughs producing over 300 publications. He is also committed to providing a strategic bridge between fundamental science and industry, he holds several patents and is engaged in a number of active industry projects in the fields biomedical diagnostics, communications and defence. He is the Director of RMIT's $50M MicroNanoResearch Facility which provides capabilities in integrated photonics, microfluidics, flexible electronics, ceramic microsystems, biomedical micro devices, sensors and nanoelectronics.

**David J. Moss** (S'83–M'88–SM'09–F'16-LF'23) is Director of the Optical Sciences Centre at Swinburne University of Technology in Melbourne, Australia, and Deputy Director of the ARC Centre of Excellence for optical microcombs for breakthrough science. He was with RMIT University in Melbourne from 2014 to 2016, the University of Sydney from 2004 to 2014, and JDSUniphase in Ottawa Canada from 1998 to 2003. From 1994 to 1998 he was with the Optical Fiber Technology Centre at Sydney University, from 1992 ro 1994 with Hitachi Central Research Laboratories in Tokyo, Japan, and from 1988 to 1992 at the National Research Council of Canada in Ottawa. He received his Ph.D. from the University of Toronto and B.Sc. from the University of Waterloo. He won the 2011 Australian Museum Eureka Science Prize and Google Australia Prize for Innovation in Computer Science. He is a Fellow of the IEEE Photonics Society, Optica (formerly the OSA), and the SPIE. His research interests include optical microcombs, integrated nonlinear optics, quantum optics, microwave photonics, ONNs, optical networks and transmission, 2D materials for nonlinear optics, optical signal processing, nanophotonics, and biomedical photonics for cancer diagnosis and therapy.